\documentclass[aps,prl,twocolumn]{revtex4-1}

\usepackage{graphicx,amsmath,amssymb,bbm}
\usepackage[utf8x]{inputenc}

\begin{document}

\title{Multiple scattering of light in superdiffusive media}

\author{Jacopo Bertolotti}
\author{Kevin Vynck}
\author{Diederik S. Wiersma}
\affiliation{European Laboratory for Non-linear Spectroscopy (LENS) \& CNR-INO,
50019
Sesto Fiorentino (Florence), Italy.}

\begin{abstract}
Light transport in superdiffusive media of finite size is studied theoretically.
The intensity Green's function for a slab geometry is found by discretizing the
fractional diffusion equation and employing the eigenfunction expansion method.
Truncated step length distributions and complex boundary conditions are
considered. The profile of a coherent backscattering cone is calculated in the
superdiffusion approximation.
\end{abstract}

\maketitle

Light transport in disordered media is characterized by a multiple scattering
process engendered by random fluctuations of the refractive index in space and
described perturbatively by expanding the field in powers of the scattering
potential~\cite{akkermansbook}. Important quantities such as the angular
distribution of backscattered and transmitted intensity can be calculated within
a diffusion approximation and the Green's function of the diffusive equation, or
diffusion propagator, be
used to describe fundamental interference effects like speckles and weak
localization~\cite{akkermansbook}. The diffusive picture has its roots in the
concept of Brownian motion, where random walkers perform independent steps of
variable length, each with finite mean and variance. By virtue of the central
limit theorem, the step length distribution after a large number of steps
approaches a Gaussian
distribution irrespectively of the microscopic transport mechanism. This rule,
however, breaks down when the probability to perform arbitrary long steps is
non-vanishing. In this case, the limiting distribution becomes a so-called
$\alpha$-stable Lévy distribution~\cite{levyoriginal} and is characteristically
heavy-tailed.
Such random walks, first studied by Mandelbrot in the framework of transport on
fractals~\cite{mandelbrotfractal}, are known as Lévy flights if steps of
arbitrary length can be performed in a unit time and Lévy walks if performed at
a finite velocity~\cite{sokolovLW}.
As transport is dominated by a few huge steps, the resulting
average distance explored by a walker increases faster with time than expected
for standard diffusion. 
This type of anomalous transport is called superdiffusion~\cite{anotrans} and has been found
to be ubiquitous in nature~\cite{ubiquity}. Superdiffusion of light has recently
been observed in heterogeneous dielectric materials ~\cite{WiersmaNature2008}
and in hot atomic vapors~\cite{kaiserlevyatom}. On the theoretical level,
superdiffusion has been modelled by employing the subordinator method
\cite{sokolowsubordination} and by generalizing the diffusion equation to
fractional order derivatives~\cite{yanovskyfractionaldiffusion}.
Previous works have evidenced the peculiar statistical properties of Lévy
motions~\cite{Chechkin2003, Koren2007} and shown that several features of real
experiments, such as properly defined boundary conditions, are nontrivial to
implement~\cite{Chechkin2003}, making the description of observable quantities
nearly impossible.

In this Letter, we develop a theoretical framework for multiple light scattering
in superdiffusive media. Our approach, which relies on the semi-analytical
solution of the fractional diffusion equation, allows to study the steady-state
transport properties of superdiffusive media while taking into account the
intrinsic finite size of actual materials and makes it possible to treat
interference effects, notably coherent backscattering, in the
``superdiffusion approximation''. In particular, we calculate the intensity
Green's function
in the superdiffusive regime for various values of the $\alpha$ coefficient and
show that arbitrary boundary
conditions and truncations in the step length distribution can be implemented.

From the microscopic point of view, the disorder averaged intensity $I$ observed
at a point $\mathbf{R}$ outside a multiple scattering medium can be written as
\begin{equation}
\begin{split}
 I(\mathbf{R}) =& \int d \mathbf{r}_1 d \, \mathbf{r}_2 \, d \mathbf{r}_3 \,
d\mathbf{r}_4 \, \phi (\mathbf{r}_1) \, \phi^{\star} (\mathbf{r}_2) \cdot \\
& \cdot f \left(\mathbf{r}_1, \mathbf{r}_2, \mathbf{r}_3, \mathbf{r}_4 \right)
\, G (\mathbf{r}_3,\mathbf{R}) \, G^{\star} (\mathbf{R},\mathbf{r}_4)
\end{split}
\label{eq:microscopicintensity}
\end{equation}
where $\phi$ is the coherent (i.e. unscattered) propagator for the amplitude
from outside the sample to the first scattering event, $G$ is the averaged
propagator for the amplitude from the last scattering event to the point of
observation $\mathbf{R}$ (i.e. the solution to the Dyson equation) and $f
\left(\mathbf{r}_1, \mathbf{r}_2, \mathbf{r}_3, \mathbf{r}_4 \right)$ is the
four-vertex propagator that contains all information about transport. When
recurrent scattering is neglected, only two two-vertex
terms contribute to $f$: the ladder term, which describes the incoherent
transport, and the most-crossed term, which leads to the coherent backscattering
cone~\cite{cbs}. More complicated combinations of these two terms can also be
used to describe speckle correlations and intensity
fluctuations~\cite{leespecklecorrelations}.

When the step length distribution $p(\ell)$ decays fast enough, the diffusion
approximation holds and the propagator for the incoherent intensity transport
satisfies the standard diffusion equation \cite{akkermansbook}. On the other
hand, for Lévy flights, the step length distribution exhibits a power-law tail
of the form
$p(\ell) \sim \ell^{-\left( \alpha +1 \right)}$ with $0 < \alpha \leq 2$, and
the
macroscopic motion is described by the diffusion-like equation \cite{anotrans}:
\begin{equation}
 \partial_t C \left(\mathbf{r},t \right) = D_{\alpha} \nabla^{\alpha} C
\left(\mathbf{r},t \right)
\label{eq:fractionaldiffusion}
\end{equation}
where $\nabla^{\alpha}$ is the symmetric Riesz fractional derivative with
respect to spatial coordinates and $D_{\alpha}$ is a generalization of the
diffusion constant.

In the normal diffusive case, finite-size effects and internal reflection at the
boundaries are handled by imposing that the propagator goes to zero at a
distance from the physical boundaries called the extrapolation
length~\cite{weitzinternalreflections}. This is possible because the Laplacian
operator is local in space and thus, the presence of boundaries does not change
the form of the propagator itself. On the other hand, the fractional Laplacian
in Eq.~\ref{eq:fractionaldiffusion} is non-local and, as such, the
superdiffusive propagator is very sensitive to the nature of the boundaries. In
fact, this propagator in non-infinite media is known only in a few particular
cases~\cite{reflectingboundarylevy}. The description of superdiffusive transport
in finite-size media therefore requires to determine the form of the propagator
for arbitrary step length distributions and boundary conditions.
\begin{figure}[tb]
	\centering
		\includegraphics[width=0.4\textwidth]{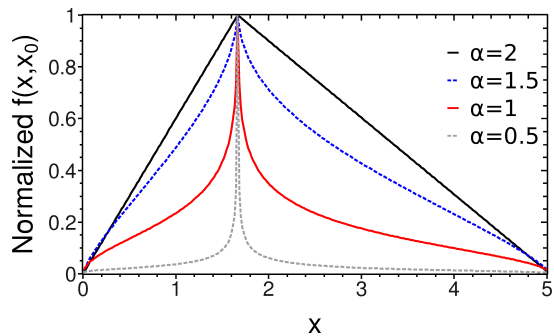}
	\caption{Normalized Green's function $f \left( x, x_0 \right)$ for a
slab with $L=5$, $x_0=L/3$, $M=300$, $D_{\alpha}=1$ and absorbing boundaries,
computed for various values of $\alpha$.}
	\label{fig:green_vs_alpha}
\end{figure}

The problem of the non-locality of the fractional Laplacian can be circumvented
by discretizing the fractional Laplacian~\cite{podlubnydiscretefractional}, i.e.
by replacing the continuous time random walk by discrete hops on a lattice and
$\nabla^{\alpha}$ by an $M \times M$ matrix, which, when applied to the vector
representing our function $C$, converges to the continuum operator when $M$ goes
to infinity. Let us consider a 1D system where space and time are discretized
(the generalization to the 3D case will be presented later in this Letter) and
define $\omega_{\left| i-j \right|}$ as the probability to perform a jump from
site $i$ to site $j$. The macroscopic transport is expected to be described by
Eq.~\ref{eq:fractionaldiffusion} when the time and space discretizations are
fine enough. Concurrently, the microscopic redistribution process that occurs at
each time interval can be written as $C_j(t_{n+1}) = \sum_i \omega_{\left| i-j
\right|} C_j (t_n)$ such that~\cite{mainardidiscretization}:
\begin{equation}
\frac{C_j (t_{n+1}) - C_j (t_n)}{\tau} = \frac{1}{\tau} \sum_{i=1}^M \left(
\omega_{\left| i-j \right|} - \delta_{i,j} \right) C_j (t_n) ,
\end{equation}
where $\delta_{i,j}$ is the Kronecker delta.

In the limit $\tau \rightarrow 0$, the left- and right-hand sides of this
equation represent $\partial_t C_j (t)$ and $D_ {\alpha} \nabla^{\alpha} C_j
(t)$, respectively. Any microscopic redistribution property, provided that it
leads to superdiffusion, can therefore be used to write a discretized version of
the fractional Laplacian. In particular, if $\Omega$ is the matrix of transition
probabilities, we have:
\begin{equation}
\label{eq:discretizedfractional}
 D_ {\alpha} \nabla^{\alpha} \sim \lim_{\tau \rightarrow 0} \frac{1}{\tau}
\left( \Omega - \mathbbm{1} \right ) .
\end{equation}
The convergence properties of this limit depend on the particular choice of the
transition probabilities. The most natural choice, $\omega_{\left| i-j \right|}
\propto \left| i-j \right|^{-\left( \alpha+1 \right)}$, is known to suffer from
a poor convergence, especially when $\alpha \rightarrow 2$ \cite{Buldyrev1992}.
A much faster convergence has been demonstrated~\cite{zoiafractional} using a
direct discretization of the fractional Laplacian, leading
to~\cite{podlubnydiscretefractional}:
\begin{equation}
 \Omega_{i,j} = \frac{1}{h^{\alpha}} \frac{\Gamma \left( -\frac{\alpha}{2} +
\left| i-j \right| \right) \Gamma \left( \alpha +1 \right)}{\pi \Gamma \left( 1+
\frac{\alpha}{2} + \left| i-j \right| \right)} ,
\end{equation}
where $h$ is the distance between two consecutive nodes on the lattice and
$\Gamma$ is the Euler gamma function. In this framework, setting $\omega_{\left|
i-j \right|}=0$ when $j$ is outside a given interval $\left[ 0,L\right]$
corresponds to the situation in which walkers stepping out of the interval
cannot ever re-enter it. Thus, reducing the infinite size matrix $\Omega$ to a
$M \times M$ matrix comes to imposing a finite size with absorbing boundary
conditions to the system~\cite{zoiafractional}.

A physical model for superdiffusion in real systems should rely on Lévy walks
rather than on Lévy flights since all jumps are bound to have a finite velocity.
The resulting spatiotemporal coupling~\cite{sokolovLW} is known to make the
description of Lévy walks difficult to handle
analytically~\cite{meerchaertlevywalk} as opposed to Lévy flights, essentially
described by Eq.~\ref{eq:fractionaldiffusion}. This coupling, however, becomes
irrelevant in the steady-state regime since the amount of time required to
perform a jump plays no role. The intensity Green's function in a 1D
superdiffusive medium for a continuous point source at $x_0$ is then given by
the following time-independent fractional differential equation:
\begin{equation}
D_{\alpha} \nabla^{\alpha} f\left( x, x_0 \right) =- \delta (x-x_0) .
\label{eq:steadylevy}
\end{equation}
Note that $f\left( x, x_0 \right)$ is the two-vertex propagator appearing in
Eq.~\ref{eq:microscopicintensity} in the superdiffusion approximation.

A complete description of the operator $\nabla^{\alpha}$ is provided by the
eigenfunctions $\psi_i$ and eigenvalues $\lambda_i$ of the matrix $\Omega -
\mathbbm{1}$, with
$h=L/M$. Then, $f$ can be expanded in terms of the eigenfunctions $\psi_i$ as a
linear combination $f\left( x \right) = \sum_i a_i \psi_i$, where $a_i$'s are
the coefficients to be determined. By substituting the above expressions in
Eq.~\ref{eq:steadylevy} and considering that $\psi_j (x) \psi_i
(x)=\delta_{i,j}$, we obtain $a_i=-\psi_i (x_0) \left(D_{\alpha}
\lambda_i\right)^{-1}$, yielding the Green's function of a 1D superdiffusive
motion of exponent $\alpha$:
\begin{equation}
 f \left( x, x_0 \right) = - \sum_{i=1}^{M} \frac{\psi_i
\left(x\right) \psi_i \left( x_0 \right)}{D_{\alpha} \lambda_i} .
\label{eq:1Dgreen}
\end{equation}
Figure~\ref{fig:green_vs_alpha} shows the normalized Green's function $f$ for
different values of $\alpha$. For $\alpha=2$ we recover the familiar triangular
shape typical of the diffusive regime while the Green's function becomes more
and more cusped when $\alpha$ decreases.

\begin{figure}[tb]
	\centering
		\includegraphics[width=0.4\textwidth]{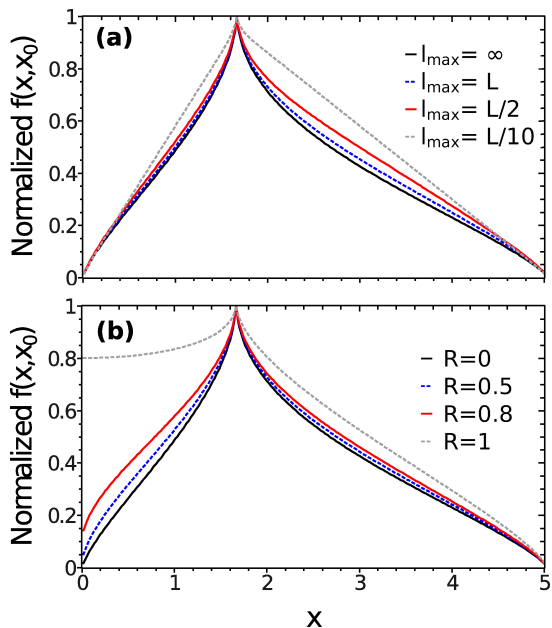}
	\caption{Normalized Green's function $f \left( x, x_0 \right)$ for a
slab with $\alpha=1.5$, $L=5$, $x_0=L/3$, $M=300$ and $D_{\alpha}=1$, computed
for various values of (a) $l_{\text{max}}$ and (b) the reflectance $R$ on the
left boundary.}
	\label{fig:green_vs_trunc_R}
\end{figure}

The microscopic transport mechanism of light in real systems is subject to
additional and more complex features, including a truncation in the step length
distribution and partially or totally reflecting boundaries. Those can be taken
into account through Eq.~\ref{eq:discretizedfractional} by modifying the
transition probabilities $\omega_{\left| i-j \right|}$ in the medium. Truncated
Lévy step distributions \cite{Mantegna1994}, which are unavoidable in
finite-size
systems~\cite{WiersmaNature2008}, can be implemented by setting to zero all
transition matrix elements where $\left| i-j \right| \ge l_{\text{max}}$ and
renormalizing $\Omega$ so that $\sum_j \omega_{\left| i-j \right|} =1 \, \forall
\, i$ before introducing the boundaries. Figure~\ref{fig:green_vs_trunc_R}a
shows
how the normalized Green's function at constant $\alpha$ changes with the
truncation length. While when $l_{\text{max}} \simeq L$ there is only a minor
correction to the shape of the Green's function,
$f \left( x, x_0 \right)$ becomes very similar
to the diffusive one when $l_{\text{max}} \ll L$.

Partially reflecting boundaries are also expected to change the shape of the
propagator. They can be implemented by considering that walkers reaching a
boundary of the superdiffusive media have a probability $R$ to be reflected. In
the case where the left boundary has reflectance $R$ while the right one is
totally absorbing, partial reflection is enforced by mapping all matrix elements
corresponding to $j<0$ onto their mirror images, yielding $\Omega_{i,j}=
\omega_{\left| i-j \right|}+ R \, \omega_{i+j+1}-\delta_{i,j}$. The case of both
partially reflecting boundaries are conceptually analogous albeit a bit more
involved due to the fact that one has to consider the possibility of performing
very long steps that might bounce forth and back a large number of times.
Figure~\ref{fig:green_vs_trunc_R}b shows the normalized Green's function for
different
values of the reflectance $R$ of the left boundary. Note that when
$R=1$ the gradient of $f \left( x, x_0 \right)$ on the left boundary goes to
zero as expected~\cite{reflectingboundarylevy}.
%
%

Up to now, we have considered the 1D case of superdiffusive
transport in finite media. Our approach can easily be extended to higher
dimensions in the case of a slab geometry. Orienting the slab such that its
interface is normal to the $x$-axis, the system becomes translationally
invariant
in both $y$- and $z$-directions. The 3D counterpart of Eq.~\ref{eq:steadylevy}
can then
be written in terms of the Fourier transform of $f(\mathbf{r},\mathbf{r}_0)$ in
the $yz$-plane as:
\begin{equation}
 D_{\alpha} \left( \nabla^{\alpha}_x - k_{\perp}^{\alpha} \right)
f \left( x, x_0, \mathbf{k}_{\perp} \right) = - \delta(x-x_0).
\label{eq:3Dlevy}
\end{equation}
Applying to Eq.~\ref{eq:3Dlevy} the approach used to find Eq.~\ref{eq:1Dgreen},
we
find:
\begin{equation}
 f (x, x_0, \mathbf{k}_{\perp} ) =  - \sum_{i=1}^{M}
\frac{\psi_i (x_0) \psi_i (x)}{D_{\alpha} \left( \lambda_i - k_{\perp}^{\alpha}
\right)} ,
\label{eq:3Dgreen}
\end{equation}
where $k_{\perp}=|\mathbf{k}_{\perp}|$. The 3D Green's function of the
superdiffusive medium can be obtained at this
point by performing an inverse Fourier transform of Eq.~\ref{eq:3Dgreen}. The
full intensity distribution in the system (including the transmission profile
\cite{WiersmaNature2008}) can be obtained upon integration over a suitable
source. As shown below, Eq.~\ref{eq:3Dgreen} can also be used to compute the
shape of the coherent backscattering cone in the superdiffusion approximation.

In the multiple scattering regime, interferences can play a major role in
determining the transport properties of the medium. In particular, in the exact
backscattered direction, the interference coming from counterpropagating paths
is always constructive as long as the system is reciprocal~\cite{akkermansbook}.
This leads to a narrow cone of enhanced albedo in reflection known as the
coherent backscattering cone. If reciprocity is not broken, the peak of the
enhanced albedo is exactly twice the common incoherent reflection and presents a
triangular cusp on the top \cite{wiersmacbs} while its exact shape depends on
transport in the medium. As a rule of thumb, long paths contribute to the
formation of the cusp and short ones to the tails of the cone. Since in a
superdiffusive regime there is no a priori reason for reciprocity to break down,
we expect longer paths to contribute more than in standard diffusion and thus,
expect a sharper peak.

The coherent component of the albedo $A$ can be calculated starting from
Eq.~\ref{eq:microscopicintensity} by setting $\mathbf{r}_1=\mathbf{r}_4$ and
$\mathbf{r}_2=\mathbf{r}_3$ ~\cite{akkermansbook}. Considering a planewave at
normal incidence on the slab interface, using the Fraunhofer approximation for
the Green's functions and assuming that the step distribution follows a
power-law, we can write:
\begin{equation*}
\begin{split} 
&\phi (\mathbf{r}_1) = x_1^{-\left( \alpha+1 \right) /2} \; e^{i
\mathbf{k}_\text{i} \cdot \mathbf{r}_1} \\
&G (\mathbf{r}_2,R) = \left( \frac{x_2}{\cos \theta} \right)^{-\left( \alpha+1
\right)/2} \; e^{-i \mathbf{k}_\text{e} \cdot \mathbf{r}_2}  \; \frac{e^{i k
R}}{4 \pi R}
\end{split}
\end{equation*}
and similarly for $\phi^{\star} (\mathbf{r}_2)$ and $G^{\star} (R,
\mathbf{r}_1)$, where $\theta$ is the angle with respect to the normal of the
slab,
$\mathbf{k}_\text{i}$ and $\mathbf{k}_\text{e}$ the wavevectors of the incident
and emergent planewaves respectively, and
$\mathbf{k}_{\perp}=(\mathbf{k}_i+\mathbf{k}_e)_{\perp}$. After substitution in
Eq.~\ref{eq:microscopicintensity} and Fourier transform, we obtain:
\begin{equation}
 A \propto \iint_{0}^{L} d x_1 d x_2 \; \left( \frac{x_1 x_2}{\cos \theta}
\right)^{-\left( \alpha+1
\right)} f (x_1,x_2, \mathbf{k}_{\perp} ) .
\end{equation}
Using the discretized approximation in Eq.~\ref{eq:3Dgreen} for the propagator
$f (x_1,x_2, \mathbf{k}_{\perp} )$,
we find the following expression for the coherent albedo from a superdiffusive
slab:
\begin{equation}
 A \propto - \sum_{x_1, x_2} \left( \frac{x_1 x_2}{\cos \theta} \right)^{-\left(
\alpha+1
\right)} \sum_{i=1}^{M} \frac{\psi_i (x_1) \psi_i (x_2)}{D_{\alpha} \left(
\lambda_i - \left| k
\sin{\theta} \right|^{\alpha} \right)} .
\end{equation}

\begin{figure}[tb]
	\centering
		\includegraphics[width=0.4\textwidth]{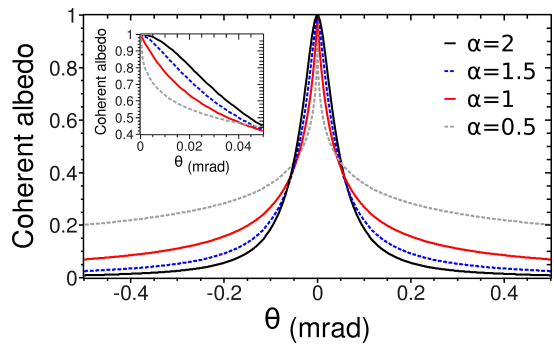}
	\caption{Normalized coherent albedo for a slab with $L=5$, $M=300$ and
$D_{\alpha}=1$, computed for various values of $\alpha$.
When $\alpha$ is decreased the contribution of long paths to the backscattering
cone is larger leading to a sharper cusp and to tails that fall down more slowly
(but still go to zero for $\theta= \pm \pi/4$).
A zoomed view of the top of the curves is shown in the inset.}
	\label{fig:cbs_vs_alpha}
\end{figure}

Figure~\ref{fig:cbs_vs_alpha} shows the normalized profile of the coherent
backscattering cone as a function of $\alpha$. When $\alpha$ is decreased the
amount of light transmitted through the sample is increased (in the
superdiffusive regime $T \propto L^{-\alpha / 2}$~\cite{Buldyrev1992}) yet, at
the same time, long steps become increasingly important, thereby making the
profile more cusped. We note that a similar effect has been predicted for
enhanced backscattering in fractal media~\cite{akkermansfractal}. Note also that
since all calculations are done considering a finite thickness, part of the
light is lost by transmission through the system, and thus, the top of the cone
for $\alpha=2$ appears
rounded~\cite{akkermansbook}.

In conclusion, we obtained a semi-analytical formulation for the intensity
Green's funtion of multiple scattered light in the superdiffusive regime
applying the eigenfunction expansion method to the discretized version of the
steady-state fractional diffusion equation. This approach makes it possible to
describe the behavior of many observable properties of superdiffusive media of
finite size with complex boundary conditions (absorbing, partially reflecting,
reflecting) as well as truncated step distributions. It also allows for the
calculation of fundamental interference effects, such as the coherent
backscattering cone, in the superdiffusion approximation.

\begin{acknowledgments}
We wish to thank Pierre Barthelemy and Stefano Lepri for useful discussion and
Igor Podlubny for pointing out relevant bibliography. We acknowledge support by
the European Network of Excellence ``Nanophotonics for Energy Efficiency'' and ENI
S.p.A. Novara.
\end{acknowledgments}

\end{document}